\documentclass[10pt]{article} 

\usepackage[utf8]{inputenc} 

\usepackage{geometry} 
\usepackage{graphicx} 

\usepackage{booktabs} 
\usepackage{array} 
\usepackage{paralist} 
\usepackage{verbatim} 
\usepackage{subfig} 
\usepackage{graphicx} 
\usepackage{fancyvrb } 
\usepackage{rotating} 
\usepackage{amssymb,amsmath} 

\usepackage{fancyhdr} 
\usepackage{sectsty}
\allsectionsfont{\sffamily\mdseries\upshape} 

\usepackage[nottoc,notlof,notlot]{tocbibind} 
\usepackage[titles,subfigure]{tocloft} 

\title{\textbf{\large On Single Qubit Quantum State Tomography}}
\author{\small Ramesh Bhandari\footnote{rbhandari@lps.umd.edu}\\[-1.0ex]
\small \textit {Laboratory for Physical Sciences, 8050 Greenmead Drive, College Park, Maryland 20740}}
\date{} 

\begin{document}
\maketitle

\begin{abstract}
In this paper, we derive analytic expressions for the starting (initial) values of the parameters of the T-matrix that is frequently employed in the construction of a theoretical density matrix in a Maximum Likelihood Estimate (MLE) procedure; this optimization procedure is used to fit the experimental data pertaining to single qubit state tomography. Appropriate starting values are critical in achieving a global minimum in the fitting process. We also indicate an analytic way of making the experimentally determined density matrix physical without resorting to the MLE process, if optimality in quantum state tomography can be disregarded, since the ultimate goal is optimality in quantum process tomography.
\end{abstract}
\thispagestyle{empty}


\section{\large Introduction}
The goal of quantum process tomography is to unravel the behavior of a device-under-test (DUT) from knowledge of the quantum states of the input qubits impinging upon the DUT as well as the output qubits. The input quantum states and the output quantum states are determined from measurements in a process called quantum state tomography. These measurements are often accompanied with experimental errors leading to violations of the required properties of the density matrix corresponding to these states. As a result, a theoretical fit to the collected experimental data must be made. Therefore, a theoretical density matrix is constructed in terms of parameters to be fitted in an optimization technique called the Maximum Likelihood Estimate (MLE), which is often employed [1]. The constructed density matrix $\rho$ must satisfy the following three requirements:\\\\
i) $\rho^\dagger=\rho$ (Hermitian)\\
ii) $ \rho$ is nonnegative definite, i.e., its eigenvalues are nonnegative (zero or greater)\\
iii) Trace {$\rho$} =1.\\\\
These three requirements of the fitted $\rho$ matrix follow from the basic definition of the $\rho$ operator: $\rho=\sum_i p_i|\psi_i><\psi_i|$ [3], where the $\psi_i$'s and the $p_i$'s define the statistical ensemble of qubits; the $\psi_i$'s are the pure states the qubit can be in and $p_i$'s are the corresponding probabilities. The first property ensures that the eigenvalues are real. The second property ensures that the probabilities are non-negative, and the third property ensures that the probabilities of the quantum state 
being in the different pure states comprising the ensemble add up to one.\\\\
In this paper, we focus on single qubit state tomography, describing the experimental determination of the density matrix, and the possible constraint violations due to experimental errors, and the subsequent necessary fit to the data in terms of a non-violating theoretical density matrix, expressed in terms of a T-matrix, whose parameters are varied in the MLE optimization process. To achieve a global minimum in the MLE process, it is then imperative to initialize the search with an appropriate set of starting values of the parameters to be fitted. These starting values are derived from the knowledge of the unphysical density matrix determined from the experimental data. In the end, we also point out an analytic solution, which provides a density matrix, which is physical but not necessarily optimal, thus circumventing the use of the MLE procedure, if optimality is not of concern.

\section{\large Construction of the Experimental Density Matrix}
We first derive theoretical expressions based on the Stokes parameters, and then relate the Stokes parameters to experimental data.\\\\
It is well known that the density matrix $\rho$ can be written as [1,3]
\begin{equation}
\rho=\frac{I+s_1\sigma_1+s_2\sigma_2+s_3\sigma_3}{2},
\end{equation}
where $I$ is a 2x2 identity matrix, $\sigma_i$'s, i=1,2,3 are the Pauli matrices and the $s_i$'s, i=1,2,3 are the normalized Stokes parameters, which are real numbers\footnote{Any 2x2 complex matrix can be written as a linear combination of the identity matrix $I$ and the three Pauli matrices, $\sigma_1, \sigma_2$, and $\sigma_3$. Noting that $Tr(\sigma_i) =0\forall{i}$ and $\sigma_i$'s (Eq. 2) are all Hermitian, the above form for $\rho$, satisfying properties i and iii (Section 1), follows immediately.}; the 2x2 Pauli spin matrices are
\begin{equation}
\sigma_1=
\left[
\begin{matrix}
0&1\\1&0
\end{matrix}
\right],
\sigma_2=
\left[
\begin{matrix}
0&-i\\i&0
\end{matrix}
\right],
\sigma_3=
\left[
\begin{matrix}
1&0\\0&-1
\end{matrix}
\right].
\end{equation}
Inserting Eq. 2 in Eq. 1, we obtain
\begin{equation}
\rho=\frac{1}{2}
\left[
\begin{matrix}
1+s_3&s_1-is_2\\s_1+is_2&1-s_3
\end{matrix}
\right].
\end{equation}
Note that properties i and iii (see Section 1) are clearly satisfied. Property ii ($det(\rho)\geq0$) leads to the requirement that 
\begin{equation}
s_1^2+s_2^2+s_3^2 \leq 1,
\end{equation}
where the equality sign holds when the density matrix $\rho$ describes a completely pure state. Eq. 4 further implies that 
\begin{equation}
s_i^2\leq 1
\end{equation}
for all i.

\subsection{\normalsize Experimental Determination of Stokes Parameters}
We immediately see from Eq. 1 that
\begin{equation}
s_i=Tr(\rho\sigma_i),
\end{equation}
where we have used the fact that $\sigma_j^2=I$ and $Tr(\sigma_j)=0, \ j=1,2,3$.
The trace of $\rho$ times an operator is the average value of the physical quantity represented by the operator. Experimentally then,
\begin{equation}
s_1=<\sigma_1> = \frac{N_D-N_A}{N_D+N_A},
\end{equation}
\begin{equation}
s_2=<\sigma_2> = \frac{N_R-N_L}{N_R+N_L},
\end{equation}
\begin{equation}
s_3=<\sigma_3> = \frac{N_H-N_V}{N_H+N_V},
\end{equation}
where the letter N with its subscript denotes the number of photon counts found upon measurement in the state corresponding to the subscript; the subscripts
D and A stand for diagonal and antidiagonal polarizations (the basis in which the matrix $\sigma_1$ is diagonal); similarly, R and L stand for right-circular and left circular polarizations (the basis in which the matrix $\sigma_2$ is diagonal), and H and V stand for horizontal and vertical polarizations that comprise the basis in which the matrix $\sigma_3$ is diagonal. Polarized photons constitute the qubits here. \\\\
The above implies 6 measurements, but use of the fact that $N_D+N_A=N_R+N_L=N_H+N_V=N$, the total number of photons (qubits) in the ensemble that are measured, reduces the above set of equations to
\begin{equation}
s_1=<\sigma_1> =2N_D/N-1,
\end{equation}
\begin{equation}
s_2=<\sigma_2> = 2N_R/N -1,
\end{equation}
\begin{equation}
s_3=<\sigma_3> = 2N_H/N-1,
\end{equation}
which correspond to 4 measurements: $N_H,N_V,N_D, N_R (N=N_H+N_V)$ that are typically made for single qubit tomography. Nevertheless, because single photon sources are not commercially available, experimentalists sometimes resort to beams of light as the source of photons to perform quantum state tomography. When that happens, the Stokes parameters are defined as
\begin{equation}
s_1=<\sigma_1> =2I_D/I-1,
\end{equation}
\begin{equation}
s_2=<\sigma_2> = 2I_R/I -1,
\end{equation}
\begin{equation}
s_3=<\sigma_3> = 2I_H/I-1,
\end{equation}
where the $I'$s stand for the measured intensities.
\\\\
From the Stokes parameters determined experimentally as above, one obtains using Eq. 3 a density matrix, which constitutes the experimentally determined density matrix. This density matrix satisfies Property i) and Property iii) listed in Section 1. The third property, Property ii) in Section 1, described via Eq. 4, may or may not be satisfied due to experimental errors. For example, it might happen that the measuring apparatus yields a value, say, for $I_H$ which is higher than the measured value of $I$, in which case the experimentally determined value of $s_3$, as given by Eq. 4, exceeds 1. Two cases arise:
\\\\
1) Eq. 4 is satisfied
\\\\
If this happens, the density matrix constructed from the measured Stokes parameters, as in Eq. 3, satisfies all the three properties mentioned in Section 1. This matrix is then the solution we are seeking. We do not need to do anything more. In other words, the MLE procedure is not required; it is redundant. 
\\\\
2) Eq. 4 is not satisfied
\\\\
When this happens, the density matrix, Eq. 3, is unphysical and must be fitted, i.e., an MLE process is necessitated. In what follows, we describe the T-matrix construction of a physical density matrix that is fitted to the experimental data in the MLE process, and analyze its properties to extract an appropriate set of starting values of the T-matrix parameters in the search of a global minimum.

\section{\large The T-Matrix and the Fitted Density Matrix}
For single qubit tomography, the fitted $\rho$ matrix is normally defined through a T-matrix in the following way (see, e.g., Refs. [1-2]):
\begin{equation}
\rho=\frac{T^\dagger T}{Tr(T^\dagger T)},
\end{equation}
where
\begin{equation}
T=
\left[
\begin{matrix}
t_1&0\\t_3+it_4&t_2
\end{matrix}
\right].
\end{equation}
$t_1, t_2, t_3, $ and $ t_4$ are real parameters, whose values are obtained in the MLE fitting process. Note that all the three requirements for the $\rho$ matrix are satisfied. This form of the T-matrix commonly appears in the literature [1-2] and was currently used in fitting single qubit tomography data; the data comprise 4 independent measurements corresponding to the 4 independent Stokes parameters, which also explains the 4 independent parameters $t_i$ used in the theoretical $\rho$ matrix Eqs. 16-17. \\\\
When the above equation for the density matrix is expanded out, we obtain
\begin{equation}
\rho=\frac{
\left[
\begin{matrix}
t_1^2+t_3^2+t_4^2&t_2(t_3-it_4)\\
t_2(t_3+it_4)&t_2^2
\end{matrix}
\right]
}
{t_1^2+t_2^2+t_3^2+t_4^2}.
\end{equation}
The matrix $\rho$ is clearly Hermitian and the trace of $\rho$ is equal to 1 as expected. A consequence of property ii, Section 1 is that
\begin{equation}
det(\rho) \geq 0,
\end{equation}
which implies for the above form, Eq. 18, that 
\begin{equation}
t_1^2t_2^2 \geq 0,
\end{equation}
(applying Eq. 19 directly to Eq. 17 also yields $|t_1t_2|\geq 0$). 
The equality sign in Eqs. 19 and 20 holds only when the density matrix corresponds to a pure quantum state. \\\\
\underline{Some Remarks}\\\\
An important observation to make is that the density matrix $\rho$ given in Eq. 18 is invariant under parameter scaling, i.e., if the $t_i$'s are scaled by a factor A, the density matrix remains unaltered \footnote{ As a result, without loss of generality, one could set $t_1^2+t_2^2+t_3^2+t_4^2=1$. But this is imposing a constraint on the free parameters $t_i$'s to be fitted, and any time one imposes a constraint such as this one, one necessarily requires a different optimization procedure like the \emph{Convex Optimization}; the maximum likelihood estimate (MLE) process does not entertain any such constraint.}. 

\subsection{\normalsize Pure Quantum State}
From Eq. 20, for a given quantum state to be a pure state, one of the following two cases must then hold :\\\\
1) $t_2=0$\\
2) $t_1=0$.
\\\\
\underline {Case 1}\\\\
Upon inserting $t_2=0$ the matrix $\rho$ necessarily reduces to 
\begin{equation}
\rho=
\left[
\begin{matrix}
1&0\\0&0
\end{matrix}
\right],
\end{equation}
which corresponds specifically to the given quantum state being a $|0>$ state: 
\begin{equation}
|0>=
\left[
\begin{matrix}
1\\0
\end{matrix}
\right].
\end{equation} 
Case 2, as we will see below, is of greater interest as it corresponds to an arbitrary pure quantum state.
\\\\
\underline{Case 2}
\\\\
Here we insert $t_1=0$. The $\rho$ matrix then reduces to
\begin{equation}
\rho=\frac{
\left[
\begin{matrix}
t_3^2+t_4^2&t_2(t_3-it_4)\\
t_2(t_3+it_4)&t_2^2
\end{matrix}
\right]
}
{t_2^2+t_3^2+t_4^2},
\end{equation}
which must also correspond to a pure state. In fact, it represents the most general form of a pure state. We see this by substituting
\begin{equation}
t_2=A\sin\theta/2,
\end{equation}
\begin{equation}
t_3=A\cos\theta/2\cos\phi,
\end{equation}
\begin{equation}
t_4=A\cos\theta/2\sin\phi,
\end{equation}
yielding
\begin{equation}
\rho=
\left[
\begin{matrix}
\cos^2\theta/2&\cos\theta/2\sin\theta/2\exp(-i\phi)\\
\cos\theta/2\sin\theta/2\exp(i\phi)&\sin^2\theta/2
\end{matrix}
\right],
\end{equation}
which corresponds to the most general representation of a (pure) normalized single qubit state:
\begin{equation}
|\psi>=\cos\theta/2|0> +\sin\theta/2\exp(i\phi)|1>,
\end{equation}
where $\theta$ and $\phi$ are the spherical polar angles representing the orientation of a single qubit on the Bloch sphere [3]; the basis state $|1>$ is given by
\begin{equation}
|1> = 
\left[
\begin{matrix}
0\\1
\end{matrix}
\right].
\end{equation}
It important to note here that because Eq. 23 corresponds to an arbitrary pure state, it must include the case of the pure state being a $|0>$ state. This is seen by inserting $\theta=0$ in Eq. 28. From Eq. 24, we then see that $t_2=0$; subsequently, inserting this value in Eq. 23 yields Eq. 21, which is nothing but Case 1 dealt with earlier.

\subsection{\normalsize Tests of the Fitted T-Matrix}
We now make the following observations with respect to the parameters of the T-matrix and their fits:\\\\
1) If it is known a priori that the quantum state under consideration is definitely a $|0>$ state, then Eq. 20 and the ensuing discussion imply that parameter $t_2$ is necessarily equal to zero. Therefore, in fitting a pure or near $|0>$ state, one may initialize $t_2$ to zero in reaching a global minimum in the MLE search for finding the optimal set of parameters. 
\\\\
2) If the quantum state under consideration is an arbitrary pure state, then parameter $t_1$ must be equal to zero, as shown in Section 3.1. Thus, in fitting the T-matrix to the experimental data, one may initialize $t_1$ to zero.
\\\\
3) Presence of non zero values for both the parameters $t_1$ and $t_2$ in the fitted T-matrix confirms that the state being measured is actually a mixed state.
\\\\
The above were some general observations that originated from the requirement of property ii of the density matrix ($det(\rho)\geq 0$) and involved assignment of specific values for the $t_1$ or the $t_2$ parameter when the quantum state under consideration was a pure state. For an arbitrary pure state (other than $|0>$), parameter $t_1$ is set to zero. But what about the assignment of initial values for the other parameters, $t_2$, $ t_3$ and $t_4$ and what about assignment in general, including for mixed states, or for that matter, any arbitrary (unknown) quantum state, which might be pure or mixed? In the next section, we show how the experimental data fix these values. These values then are used as starting values in the search of the global minimum. 


\subsection{\normalsize Starting Values of the $t_i$'s in the MLE Search}
In order for the MLE process to yield a global minimum (instead of a local minimum), it is imperative to assign good starting values for the $t_i$'s. In this section, we provide explicit expressions for the initial starting values. Our approach is one of relating the $t_i$'s to the measured (normalized) Stokes parameters, $s_1, s_2,$ and $s_3$. 
Comparison of Eq. 3 with Eq. 18 then yields
\begin{equation}
s_3=\frac{t_1^2+t_3^2+t_4^2-t_2^2}{t_1^2+t_2^2+t_3^2+t_4^2},
\end{equation}
\begin{equation}
s_1=\frac{2t_2t_3}{t_1^2+t_2^2+t_3^2+t_4^2},
\end{equation}
\begin{equation}
s_2=\frac{2t_2t_4}{t_1^2+t_2^2+t_3^2+t_4^2}.
\end{equation}
Solving for the $t_i$'s, one obtains after some algebra, 
\begin{equation}
t_1^2=\frac{(1-s_3^2-s_1^2-s_2^2)}{(1-s_3)^2}t_2^2,
\end{equation}
\begin{equation}
t_3=(\frac{s_1}{1-s_3})t_2,
\end{equation}
\begin{equation}
t_4=(\frac{s_2}{1-s_3})t_2 .
\end{equation}
Eqs. 33, 34, and 35 give expressions for $t_1, t_3$, and $t_4$ in terms of the Stokes parameters, $s_1, s_2$, and $s_3$, and the parameter $t_2$.

\subsubsection{Verification of the Derived Expressions}
1) From Eq. 3, 
\begin{equation}
det(\rho)=\frac{t_1^2t_2^2}{(t_1^2+t_2^2+t_3^2+t_4^2)^2},
\end{equation}
which, upon substitution of Eqs. 33, 34, and 35, reduces to 
\begin{equation}
det(\rho)=(1-s_1^2-s_2^2-s_3^2)/4
\end{equation}
in agreement with the direct calculation of the determinant of $\rho$ from Eq. 3.
\\\\
2) For an arbitrary pure state, $1-s_1^2-s_2^2-s_3^2=0$, which implies from Eq. 33 that $t_1=0$, as ascertained earlier. The case $s_3=1$ is treated later.
\\\\
3) Considering pure states further, if parameter $s_1$ is set equal to 1 (which implies $s_2=s_3=0$), then $t_3=t_2$ (from Eq. 34) and $t_4=0$ (from Eq. 35) and $t_1=0$ (from Eq. 33). Substituting in Eq. 18, one obtains 
\begin{equation}
\rho=\frac{1}{2}
\left[
\begin{matrix}
1&1\\1&1
\end{matrix}
\right],
\end{equation}
which is what one would obtain by inserting $s_1=1$, and $s_2=s_3=0$ in Eq. 3. Note that the density matrix in Eq. 38 corresponds to the quantum state $|D>=\frac{1}{\sqrt{2}}(|0>+|1>)$.
\\\\
4) Similarly, if $s_2=1$ (which implies $s_1=s_3=0$), one finds from Eqs. 33, 34, and 35 that $t_4=t_2$, $t_3=0$, and $t_1=0$. Upon insertion in Eq. 18, one obtains 
\begin{equation}
\rho=\frac{1}{2}
\left[
\begin{matrix}
1&-i\\i&1
\end{matrix}
\right],
\end{equation}
which is what one would obtain by inserting $s_2=1$, and $s_1=s_3=0$ in Eq. 3. Note that Eq. 39 corresponds to the quantum state $|R>=\frac{1}{\sqrt{2}}(|0>+i|1>)$.
\\\\
5) What about the $s_3=1$ case?
\\\\
Here we invert Eq. 33 to express parameter $t_2$ in terms of parameter $t_1$:
\begin{equation}
t_2^2=\frac{(1-s_3)^2t_1^2}{1-s_3^2-s_1^2-s_2^2}.
\end{equation}
As $s_3 \rightarrow 1$, parameters $s_1$ and $s_2$ each approach zero. Subsequently, simplifying the fraction on the right-hand-side, or equivalently applying L'Hospital's rule, $t_2$ approaches zero, yielding a density matrix of the form Eq. 21. This is just Case 1 in Section 3.1, as one would expect.
\\\\
6) Completely Mixed State: here $s_1=s_2=s_3=0$. Inserting these values in Eqs. 33-35 yields $t_1=t_2$ and $t_3=t_4=0$. Inserting them in Eq. 18 gives
\begin{equation}
\rho=\frac{1}{2}
\left[
\begin{matrix}
1&0\\0&1
\end{matrix}
\right],
\end{equation}
which is the density matrix corresponding to a completely mixed state.
\\\\
Thus, verification of Cases 1-6 validates the derived Eqs. 33-35 and Eq. 40 (we can derive and verify other special cases involving, e.g., negative Stokes parameters, but we leave those to the reader).

\section{\large Conclusions}
1) If the experimental data satisfy Eq. 4, the density matrix constructed from the measured Stokes parameters, as in Eq. 3, satisfies all the three properties mentioned in Section 1. This matrix is then the solution we are seeking. We do not need to employ the MLE procedure; it is not required; it is redundant. 
\\\\
2) If Eq. 4 is violated, the density matrix, Eq. 3, is unphysical and must be fitted, i.e., an MLE process may be employed, with Eqs. 33-35 constituting the initial starting values of the T-matrix parameters in the MLE procedure. Due to the violation of the constraint, Eq. 4, the numerator in Eq. 33 becomes negative, invalidating Eq. 33. In this situation, we simply modify the numerator to equal zero. In using Eqs. 33-35 as starting values in the global minimum search, we may fix the value of parameter $t_2$ to 1.
\\\\
If the experimentally measured value of $s_3\le 1-\epsilon$, where $\epsilon$ is a small number determined from some fitting experimentation, e.g., $=10^{-1}$), basically corresponding to a predominantly $|0>$ quantum state, then set $t_2=0$, with parameters $t_1$, $t_3$, and $t_4$ assigned arbitrary non-zero values; one may, for example, set these equal to 1 each.
\\\\
\underline{An Interesting Observation}
\\\\
Clearly, if we were to redefine $s_1^2$, $s_2^2$, and $s_3^2$ by dividing each one of them by their sum ($s_1^2+s_2^2+s_3^2$), then Eq. 4 is satisfied, and Eq. 3 becomes physical, and is a valid solution. However, it may not necessarily correspond to a global minimum solution for which an MLE procedure is required. The physical interpretation of this solution is that it corresponds to a quantum state that has been scaled back to lie on the Bloch (or Poincare) sphere, without change in its direction. In the absence of the availability of the MLE software, or because of of time-constraints, this solution can serve as a valid, physical solution en route to the determination of the process matrix, which is the ultimate goal in quantum tomography.


\section{\large Summary}
We have provided a description of the construction of the density matrix of an unknown quantum state, in a process called quantum state tomography. A density matrix is first constructed from the collected experimental data, which, more often than not, is accompanied by errors. As a result, the constructed density matrix may not always be physical. To alleviate this problem, a theoretical physical density matrix is subsequently constructed in terms of what is called a T-matrix in literature [1,2]. The parameters of this matrix are then fitted to the experimental data in a process called the MLE optimization technique to obtain an appropriate physical density matrix. In this paper, we analyze the mathematical form of the T-matrix in detail, which then yields appropriate initialization (starting values) of the parameters to be fitted in order to obtain the optimal set in the MLE fitting process. It is interesting to note that the MLE process is, in fact, not required if the initial experimentally determined density matrix is determined to meet all the required properties of a physical density matrix, i.e., the experimentally constructed density matrix is the same as any fitted solution. Furthermore, we show how the data can be tweaked to yield a physical density matrix without going through the MLE process. 
\section{\large Acknowledgment}
The author thanks Nick Peters for useful discussions on the experimental aspects of quantum tomography.


\end{document}